\documentclass[twocolumn,amssymb,amsmath,aps,pra,nobibnotes,noshowpacs]{revtex4}
\usepackage{graphicx}
\usepackage{docs}

\begin{document}

\title{Generation of Pulsed Polarization Entangled Two-Photon
State \\via Temporal and Spectral Engineering}

\author{Yoon-Ho Kim} \email{kimy@ornl.gov}
\affiliation{Center for Engineering Science Advanced Research\\ Computer
Science and Mathematics Division\\ Oak Ridge National Laboratory\\ Oak Ridge,
Tennessee 37831, U.S.A.\\Tel:865-241-0912; Fax:865-241-0381}

\author{Warren P. Grice}
\affiliation{Center for Engineering Science Advanced Research\\ Computer
Science and Mathematics Division\\ Oak Ridge National Laboratory\\ Oak Ridge,
Tennessee 37831, U.S.A.\\Tel:865-241-0912; Fax:865-241-0381}

\date[To appear in Journal of Modern Optics]{}

\begin{abstract}
The quantum state of the photon pair generated from type-II spontaneous
parametric down-conversion pumped by a ultrafast laser pulse exhibits strong
decoherence in its polarization entanglement, an effect which can be attributed
to the clock effect of the pump pulse or, equivalently, to distinguishing
spectral information in the two-photon state. Here, we propose novel temporal
and spectral engineering techniques to eliminate these detrimental decoherence
effects. The temporal engineering of the two-photon wavefunction results in a
universal Bell-state synthesizer that is independent of the choice of pump
source, crystal parameters, wavelengths of the interacting photons, and the
bandwidth of the spectral filter. In the spectral engineering technique, the
distinguishing spectral features of the two-photon state are eliminated through
modifications to the two-photon source. In addition, spectral engineering also
provides a means for the generation of polarization-entangled states with novel
spectral characteristics: the frequency-correlated state and the
frequency-uncorrelated state.
\end{abstract}


\maketitle

\section{Introduction}\label{intro}

Quantum entanglement \cite{eprb}, once discussed only in the context of the
foundations of quantum mechanics, is now at the heart of the rapidly developing
field of quantum information science \cite{nielson}. Many researchers are
hoping to exploit the unique features of the entangled states in order to
surpass the ``classical limit'' in applications such as quantum lithography
\cite{litho}, the quantum optical gyroscope \cite{gyro}, quantum clock
synchronization and positioning \cite{clock}, etc.

A particularly convenient and reliable source of entangled particles is the
process of type-II spontaneous parametric down-conversion (SPDC)
\cite{klyshkobook,kiess,shih1,rubin,kwiat1}, in which an incident pump photon
is split into two orthogonally polarized lower energy daughter photons inside a
crystal with a $\chi^{(2)}$ nonlinearity. Initially, the photon pairs are
entangled in energy, time, and momentum due to the energy and momentum
conservation conditions that govern the process. In addition, polarization
entanglement may be obtained as a result of specific local operations on the
photon pair \cite{kiess,shih1,rubin,kwiat1}. An optical process such as this
has the advantage that the photons, once generated, interact rather weakly with
the environment, thus making it possible to maintain entanglement for
relatively long periods of time.

In this paper, we propose and analyze efficient generation schemes for pulsed
two-photon polarization entangled states. Through temporal or spectral
engineering of the two-photon state produced in the type-II SPDC process pumped
by a ultrafast laser pulse (ultrafast type-II SPDC), it is possible to generate
pulsed polarization entangled states that are free of any post-selection
assumptions. Pulsed polarization entangled states are an essential ingredient
in many experiments in quantum optics. They are useful, for example, as
building blocks for entangled states of three or more photons \cite{keller}.
(In general, the SPDC process only results in two-photon entanglement.) Pulsed
two-photon entangled states are also useful in practical quantum cryptography
systems, since the well-known arrival times permit gated detection.

We begin in section \ref{type2spdc} with a discussion of the limitations of the
state-of-the-art techniques for the generation of polarization-entangled photon
pairs. In section \ref{temporal}, we present a universal Bell-state
synthesizer, which makes use of a novel interferometric method to temporally
engineer the two-photon wavefunction. We follow up in section \ref{spectral}
with an analysis of the spectral properties of the two-photon wavefunction
produced in ultrafast type-II SPDC. We then propose a method for the efficient
generation of pulsed polarization entanglement via the spectral engineering of
the two-photon wavefunction. We also discuss two two-photon polarization
entangled states with novel spectral characteristics: the frequency-correlated
state; and the frequency-uncorrelated state. Entangled states with such
spectral properties might be useful for quantum-enhanced positioning and in
multi-source interference experiments.

\section{Two-Photon Entanglement in ultrafast type-II SPDC}\label{type2spdc}

\begin{figure}[htb]
\includegraphics[width=3.in]{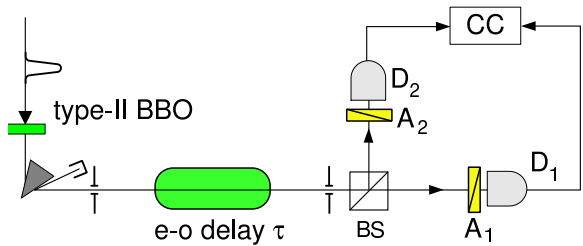}
\caption{\label{fig:type2} Typical experimental setup for preparing a
Bell-state using collinear type-II SPDC. Here, two out of four possible
biphoton amplitudes are post-selected by the coincidence circuit, see
Ref.~\cite{shih1,rubin}. One can also utilize noncollinear type-II SPDC in
which the amplitude post-selection assumption is not necessary, see
Ref.~\cite{kwiat1}. }
\end{figure}

Let us briefly review one of the standard techniques for generating
polarization entangled two-photon state via type-II SPDC
\cite{shih1,rubin,kwiat1}. A typical experimental setup is shown in
Fig.~\ref{fig:type2}. A type-II nonlinear crystal is pumped with a UV laser
beam and the orthogonally polarized signal and idler photons, which are in the
near infrared, travel collinearly with the pump. After passing through a prism
sequence to remove the pump, the signal-idler photon pair is passed through a
quartz delay circuit (e-o delay $\tau$) before the beamsplitter (BS) splits the
SPDC beam into two spatial modes. A polarization analyzer and a single-photon
detector are placed at each output port of the beamsplitter for polarization
correlation measurements. In this case, the two quantum mechanical amplitudes
in which both photons end up at the same detector are not registered since only
coincidence events are considered. That is, a state post-selection has been
made \cite{shih1,rubin,decaro}. A noncollinear type-II SPDC method developed
later resolved this state post-selection problem and is usually regarded as the
``standard'' method for generating polarization entangled photon pairs
\cite{kwiat1}.

Both the collinear and noncollinear type-II SPDC methods work very well for the
generation of polarization entangled photon pairs when the UV pump laser is
continuous wave (cw). In this case, however, there is no information available
regarding the photons' arrival times at the detectors. Such timing information
can be quite useful in certain applications, as discussed in section
\ref{intro}. If the UV pump has the form of an ultrafast optical pulse
($\approx 100$ fsec), then the photon pair arrival times can be known within a
time interval on the order of the pump pulse duration. However, it has been
theoretically and experimentally shown that, in general, type-II SPDC suffers
the loss of quantum interference if the pump is delivered in the form of an
ultrafast pulse \cite{pulsedspdctheory,pulsedspdcexp}.

We will first briefly review the theoretical treatment of the ultrafast type-II
SPDC process and discuss the physical mechanism of the loss of quantum
interference (or decoherence). With this understanding in hand, we then discuss
in the subsequent sections two methods for eliminating this decoherence through
spectral and temporal engineering of the two-photon state.

From first-order perturbation theory, the quantum state of type-II SPDC may be
expressed as \cite{rubin,pulsedspdctheory}
\begin{equation}
|\psi\rangle = -\frac{i}{\hbar}\int_{-\infty}^\infty dt \, \mathcal{H}
|0\rangle,\label{state1}
\end{equation}
where
\[
\mathcal{H}=\epsilon_0 \int d^3 \vec{r} \, \chi^{(2)} E_p(z,t) E_o^{(-)}
E_e^{(-)}
\]
is the Hamiltonian governing the SPDC process. The pump electric field,
$E_p(z,t)$, is considered classical and is assumed to have a Gaussian shape in
the direction of propagation. The operator $E_o^{(-)}$ ($E_e^{(-)}$) is the
negative frequency part of the quantized electric field of o-polarized
(e-polarized) photon inside the crystal. Integrating over the length of the
crystal, $L$, Eq.~(\ref{state1}) can be written as
\begin{equation}
|\psi\rangle=C \iint d\omega_e d\omega_o \, {\rm sinc}\left(\frac{\Delta
L}{2}\right) \mathcal{E}_p(\omega_e+\omega_o) a_e^\dagger(\omega_e)
a_o^\dagger(\omega_o)|0\rangle,\label{spdcstate}
\end{equation}
where $C$ is a constant and $\Delta\equiv
k_p(\omega_p)-k_o(\omega_o)-k_e(\omega_e)$. The pump pulse is described by
$\mathcal{E}_p(\omega_e+\omega_o)=\exp\{-(\omega_e+\omega_o-\Omega_p)^2/\sigma_p
^2\}$, where $\sigma_p$ and $\Omega_p$ are the bandwidth and the central
frequency of the pump pulse, respectively.

\begin{figure*}[htb]
\includegraphics[width=5.5in]{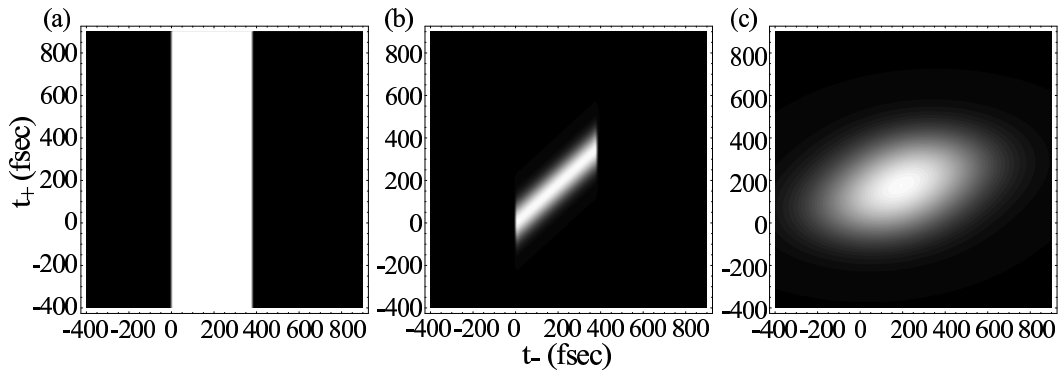}
\caption{\label{fig:wpacket} Calculated two-photon wavefunction $\Pi(t_+,t_-)$
for type-II SPDC. (a) For a cw-pumped case. The two-photon wavefunction is
independent of $t_+$ and has the rectangular shape in $t_-$. (b) For a 100 fsec
pump pulse. It is strongly asymmetric. (c) With 5 nm bandwidth spectral
filters. The two-photon wavefunction is expanded.}
\end{figure*}

For the experimental set-up shown in Fig.~\ref{fig:type2}, the electric field
operators at the detectors are
\begin{widetext}
\begin{eqnarray}
E_1^{(+)}&=&\frac{1}{\sqrt{2}} \int d\omega' \, \{\cos\theta_1 \,
e^{-i\omega'(t_1+\tau)}a_e(\omega')-\sin\theta_1 \,
e^{-i\omega't_1}a_o(\omega')
\},\nonumber\\
E_2^{(+)}&=&\frac{i}{\sqrt{2}} \int d\omega' \, \{\cos\theta_2 \,
e^{-i\omega'(t_2+\tau)}a_e(\omega')+\sin\theta_2 \,
e^{-i\omega't_2}a_o(\omega') \},\nonumber
\end{eqnarray}
\end{widetext}
where $\theta_1$ and $\theta_2$ are the angles of the polarization analyzers
$A_1$ and $A_2$ and $\tau$ is the e-o delay. Here we have assumed no spectral
filtering before detection. The coincidence count rate at the detectors is
proportional to
\begin{eqnarray}
R_c &\propto& \int dt_1 \int dt_2 \, |\langle 0 | E_2^{(+)} E_1^{(+)} | \psi
\rangle |^2 \nonumber\\&=&  \int dt_+ \int dt_- \,
|\mathcal{A}(t_+,t_-)|^2,\label{coinc}
\end{eqnarray}
where $t_+=(t_1+t_2)/2$ and $t_-=t_1-t_2$.

The two-photon amplitude $\mathcal{A}(t_+,t_-)$ has the form
\begin{widetext}
\begin{equation}
\mathcal{A}(t_+,t_-) = \cos\theta_1\sin\theta_2\Pi(t_+,t_-+\tau) -
\sin\theta_1\cos\theta_2\Pi(t_+,-t_-+\tau),\label{biphoton}
\end{equation}
where
\begin{eqnarray}
\Pi(t_+,t_-) =
    \left\{
        \begin{array}{ll}
        e^{-i\Omega_p t_+} e^{-\sigma_p^2\{t_+ - [D_+/D] t_-\}^2} &
\text{for~~} 0<t_-<DL \\
        0 & \text{otherwise.}
        \end{array}
    \right.\label{pifn}
\end{eqnarray}
\end{widetext}
The parameters $D_+$ and $D$ are defined to be
\begin{eqnarray}
D_+ &=&\frac{1}{2}\left(\frac{1}{u_o(\Omega_o)}+\frac{1}{u_e(\Omega_e)}\right)-
\frac{1}{u_p(\Omega_p)}, \nonumber\\
D&=&\frac{1}{u_o(\Omega_o)}-\frac{1}{u_e(\Omega_e)},\nonumber
\end{eqnarray}
where, for example, $u_o(\Omega_o)$ is the group velocity of o-polarized photon
of frequency $\Omega_o$ in the crystal.

It is clear from Eqs.~(\ref{coinc}) and~(\ref{biphoton}) that the degree of
quantum interference is directly related to the amount of overlap between the
two two-photon wavefunctions $\Pi(t_+,t_-+\tau)$ and $\Pi(t_+,-t_-+\tau)$.
Figure~\ref{fig:wpacket} shows calculated two-photon wavefunctions for type-II
SPDC in a 2 mm Beta-Barium Borate (BBO) crystal with a central pump wavelength
of 400 nm. The two-photon wavefunction for the cw-pumped case is symmetric in
$t_+$ and $t_-$. As shown in Fig.~\ref{fig:wpacket}(a), it has a rectangular
shape in the $t_-$ direction and extends to infinity in the $t_+$ direction. In
the case of ultrafast type-II SPDC, the two-photon wavefunction is strongly
asymmetric, see Fig.~\ref{fig:wpacket}(b). As we shall show shortly, this is
the origin of loss of quantum interference in ultrafast type-II SPDC.

\begin{figure}[htb]
\includegraphics[width=2.4in]{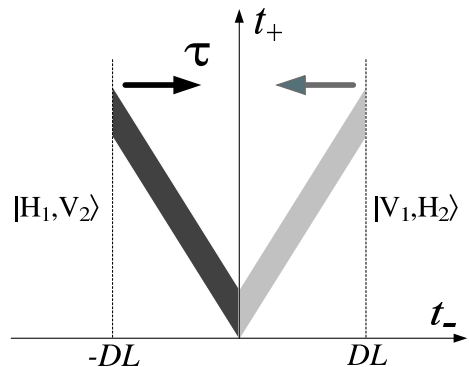}
\caption{\label{fig:overlap} Distribution of two-photon amplitude
$\mathcal{A}(t_+,t_-)$ for Eq.~(\ref{biphoton}). Due to little overlap between
$\Pi(t_+,t_-+\tau)$ and $\Pi(t_+,-t_-+\tau)$, quantum interference visibility
cannot be high. In the cw-pumped case, however, 100\% quantum interference
should occur, see Fig.~\ref{fig:wpacket}(a), if $\tau$ is correctly chosen. }
\end{figure}

Having learned the exact shape of the two-photon wavefunction $\Pi(t_+,t_-)$,
we are now in a position to study the overlap of the wavefunctions in
Eq.~(\ref{biphoton}). The situation is well illustrated in
Fig.~\ref{fig:overlap}. When $\tau=0$, there is no overlap at all and hence no
quantum interference. Increasing $\tau$ brings the two two-photon wavefunctions
together and they begin to overlap. This increases the indistinguishability of
the two wavefunctions, which represent the two-photon polarization states
$|V_1,H_2\rangle$ and $|H_1,V_2\rangle$. Because the two wavefunctions are
symmetric in cw-pumped type-II SPDC, perfect overlap may be achieved with the
proper value of $\tau$, see Fig.~\ref{fig:wpacket}(a). With an ultrafast pump,
however, the wavefunctions are asymmetric and perfect overlap is not possible.
The amount of relative overlap may be increased in one of two ways: (i) by
decreasing the thickness of the crystal; or (ii) through the use of narrowband
spectral filters to expand the two-photon wavefunction, see
Fig.~\ref{fig:wpacket}(c). Both methods increase the relative amount of
overlap, which in turn increase the overall indistinguishability of the system.
The drawback of these methods, obviously, is the reduced number of available
entangled photon pairs.

\begin{figure}[tb]
\includegraphics[width=3.0in]{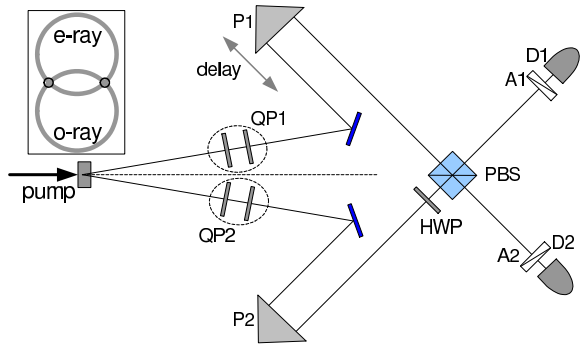}
\caption{\label{fig:setup}Universal Bell-state synthesizer. Non-collinear
type-II SPDC is used to prepare an initial polarization state which is in a
mixed state. QP1 and QP2 are thin quartz plates with optic axes oriented
vertically and HWP is the $\lambda/2$ plate oriented at $45^\circ$. }
\end{figure}

\section{Temporal Engineering of the Two-Photon State}\label{temporal}

In this section, we present a method in which complete overlap of two
two-photon wavefunctions may be achieved with no change in the spectral
properties of the photons. Consider the experimental setup shown in
Fig.~\ref{fig:setup}. A type-II BBO crystal is pumped either by a cw- or by a
pulsed- pump laser. As in Ref.~\cite{kwiat1}, we restrict our attention to the
photons found in the intersections of the cones made by the e- and o-rays
exiting the crystal \cite{note}. A $\lambda/2$ plate rotates the polarization
in one arm before the two photons are brought together at a polarizing beam
splitter (PBS). Note that since the two photons always have the same
polarizations when they reach the PBS, they always exit different ports. Thus,
the state post-selection assumption is not necessary.

\begin{figure}[thp]
\includegraphics[width=3in]{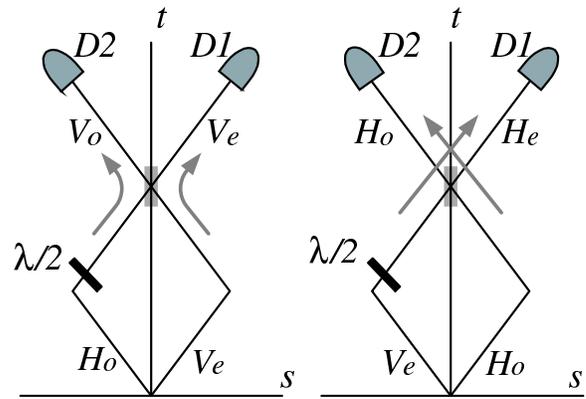}
\caption{\label{fig:feynman}Feynman alternatives for the experimental setup
shown in Fig.~\ref{fig:setup}. Note that e-ray (o-ray) of the crystal is always
detected by $D_1$ ($D_2$) and there are only two two-photon alternatives
(reflected-reflected and transmitted-transmitted). If the path length
difference between the two arms of the interferometer is zero, the two
alternatives (r-r and t-t) are indistinguishable in time regardless of the
choice of the pump source, crystal thickness, and spectral filtering.}
\end{figure}

The two Feynman alternatives (reflected-reflected and transmitted-transmitted)
leading to coincidence detection are shown in Fig.~\ref{fig:feynman}. Note
that there are only two two-photon amplitudes and in both cases the e-ray
(o-ray) of the crystal is always detected by $D_1$ ($D_2$). This means that,
unlike the experiment discussed in the previous section, any spectral or
temporal
differences between the o- and e-photons provides no information at the
detectors that might make it possible to distinguish between the two
wavefunctions. The two biphoton wavefunctions, which represent
$|H_1\rangle|H_2\rangle$ and $|V_1\rangle|V_2\rangle$, are therefore quantum
mechanically indistinguishable so that the state exiting the PBS is
$$
|\Phi\rangle = \frac{1}{\sqrt{2}}\left(|H_1\rangle|H_2\rangle + e^{i\varphi}
|V_1\rangle|V_2\rangle\right),
$$
where $\varphi$ is the phase between the two terms. This phase may be adjusted
by tilting the quartz plates QP1 and QP2. By setting $\varphi=0$, the Bell
state
$$
|\Phi^{(+)}\rangle = \frac{1}{\sqrt{2}}\left(|H_1\rangle|H_2\rangle +
|V_1\rangle|V_2\rangle\right)
$$
is attained. From this state, only simple linear operations are needed to
transform to any of the other Bell states.

Let us now analyze this interferometer more formally. If we assume that the
phase difference between the two amplitudes $\varphi$ is set to zero, then the
electric field operators that reach the detectors may be written as
\begin{widetext}
\begin{eqnarray}
E_1^{(+)}&=&\int d\omega' \, \{\cos\theta_1 \,
e^{-i\omega'(t_1+\tau)}a_{Ve}(\omega')-\sin\theta_1 \,
e^{-i\omega't_1}a_{He}(\omega')
\},\nonumber\\
E_2^{(+)}&=&\int d\omega' \, \{\cos\theta_2 \,
e^{-i\omega't_2}a_{Vo}(\omega')-\sin\theta_2 \,
e^{-i\omega'(t_2+\tau)}a_{Ho}(\omega') \},\nonumber
\end{eqnarray}
\end{widetext}
where, for example, $a_{Vo}(\omega')$ is the annihilation operator for a photon
of frequency $\omega'$ with vertical polarization which was originally created
as the o-ray of the crystal.

\begin{figure}[tb]
\includegraphics[width=3.0in]{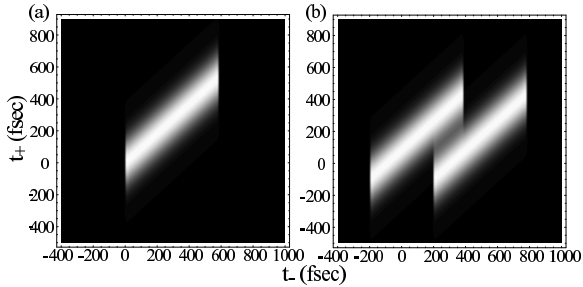}
\caption{\label{fig:temporal} Temporal engineered two-photon amplitude. This
figure shows how the two two-photon wavefunctions behave as $\tau$ is
introduced. (a) Two two-photon wavefunctions overlap completely when $\tau=0$
(compare it with Fig.~\ref{fig:overlap}). (b) When $\tau=200$ fsec, there is
almost no overlap.}
\end{figure}

Then the two-photon amplitude $\mathcal{A}(t_+,t_-)$ may be expressed as
\begin{widetext}
\begin{equation}
\mathcal{A}(t_+,t_-)=\cos\theta_1\cos\theta_2\Pi(t_+ + \tau/2, t_-+\tau) +
\sin\theta_1\sin\theta_2 \Pi(t_+ + \tau/2, t_- - \tau).\label{biphoton2}
\end{equation}
\end{widetext}
In contrast to the amplitude represented by Eq.~(\ref{biphoton}), the
two-photon wavefunctions on the right-hand side of Eq.~(\ref{biphoton2})
overlap completely when $\tau=0$, regardless of their shape, see
Fig.~\ref{fig:temporal}. Since complete overlap is possible for any shape of
the two-photon wavefunction, this method should work for both cw-pumped and
pulse-pumped SPDC. As such, this interferometer may be considered as a
universal Bell-state synthesizer $-$ perfect quantum interference may be
observed regardless of pump bandwidth, crystal thickness, or SPDC wavelength,
with no need for spectral filters.

\begin{figure}[htb]
\includegraphics[width=2.5in]{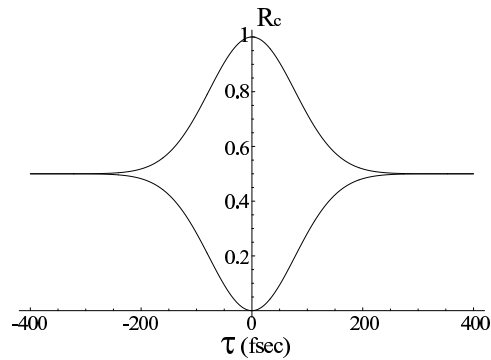}
\caption{\label{fig:curve} Calculated space-time interference pattern as a
function of delay $\tau$ in fsec. Upper line is for
$\theta_1=\theta_2=45^\circ$ and lower line is for
$\theta_1=-\theta_2=45^\circ$. The calculation is done for a 3 mm thick type-II
BBO crystal pumped by a 400 nm pump pulse with bandwidth approximately 2 nm. }
\end{figure}

The typical peak-dip effect may be observed by varying the delay $\tau$, i.e.,
$$
R_c =
    \left\{
        \begin{array}{ll}
        \frac{1}{2} + \frac{1}{2}\exp(-D_+^2 \sigma_p^2 \tau^2/2D^2) &
\text{for~~} \theta_1=\theta_2=45^\circ \\
        \frac{1}{2} - \frac{1}{2}\exp(-D_+^2 \sigma_p^2 \tau^2/2D^2) &
\text{for~~}
        \theta_1=-\theta_2=45^\circ.
        \end{array}
    \right.
$$
Note that the thickness of the crystal $L$ does not appear in the above
equation and all other experimental parameters, $D_+$, $\sigma_p$, and $D$,
only affect the width of the quantum interference, not the maximum visibility.
Figure~\ref{fig:curve} shows the calculated coincidence rate as a function of
the delay $\tau$. It is easy to see that the coincidence rate $R_c$ is a
monotonically varying function of $\tau$, with complete quantum interference
expected at $\tau=0$. This insensitivity to small changes in path length
difference provides for a stable source of high-quality Bell-states. This
robustness is not found in two-crystal or double-pulse pump type-II SPDC
schemes, where a small phase change in the interferometer results in sinusoidal
modulations at the pump central wavelength. Such schemes require active phase
stabilization and, therefore, are not practical \cite{kim3}. Our scheme does
not suffer this disadvantage.

In addition, if $\tau\neq 0$, the amplitudes do not overlap completely and a
more general state is generated:
$$
\rho = \varepsilon \rho_{ent} + (1-\varepsilon)\rho_{mix},
$$
where $\rho_{ent}=|\Phi^{(+)}\rangle\langle\Phi^{(+)}|$ and $0 \leq \varepsilon
\leq 1$. Such partially entangled states are called Werner states and have
recently been the subject of experimental studies in the cw domain
\cite{werner}. Our scheme provides an efficient way to access a broad range of
two-qubit states in both the pulsed and cw domains.

\section{Spectral Engineering of the Two-Photon State}\label{spectral}

Using the temporal engineering technique discussed in the preceding section, it
is possible to generate polarization entangled states with either cw or pulsed
pumping schemes. It should be pointed out that neither the asymmetry of the
two-photon wavefunction nor the spectral properties of the photon pairs are
affected by this technique. In this section, we are interested in removing the
asymmetry shown in Fig.~\ref{fig:wpacket}(b) through a careful choice of
wavelengths of photons involved in the interaction. In this way, it is possible
to obtain pulsed polarization-entangled photon pairs directly.

\begin{figure}[t]
\includegraphics[width=2in]{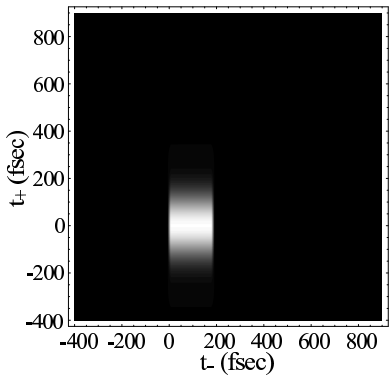}
\caption{\label{fig:symmetry} Spectrally engineered two-photon wavefunction.
Note that the asymmetry shown in Fig.~\ref{fig:wpacket}(b) has disappeared. See
text for details.}
\end{figure}

Recall that the source of poor wavefunction overlap in ultrafast type-II SPDC
is the asymmetry introduced by a coupling between $t_+$ and $t_-$ in the
two-photon wavefunction given in Eq.~(\ref{pifn}). The approach here is to
remove the coupling term so that the two-photon wavefunction becomes symmetric,
as originally proposed by Keller and Rubin \cite{pulsedspdctheory}. An example
of such a wavefunction is shown in Fig.~\ref{fig:symmetry}. Note that it has a
rectangular shape in $t_-$ (just as in cw-pumped type-II SPDC) and has a
Gaussian shape in $t_+$ due to the Gaussian pulse envelope. It is not difficult
to see that two-photon wavefunctions of this type may be completely overlapped,
thus yielding full quantum interference. Therefore, pulsed
polarization-entangled photon pairs may be generated using the well-known
techniques described in Ref.~\cite{shih1,kwiat1}. It is also expected that a
triangular shaped correlation function, which has been observed in cw-pumping
SPDC, should be observed in ultrafast type-II SPDC, as well.

\begin{figure}[htb]
\includegraphics[width=2.5in]{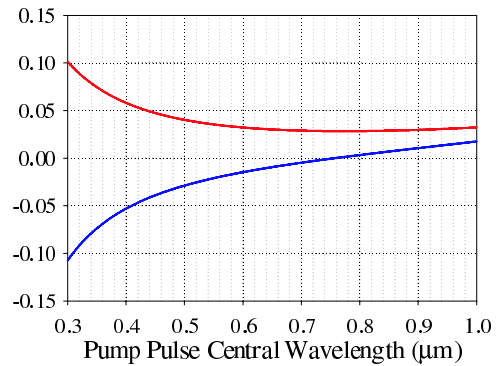}
\caption{\label{fig:dpdm} $D_+$ (lower) and $D$ (upper), shown in units of
$1/c$, for a type-II BBO crystal. Note that $D_+=0$ at the pump wavelength of
757 nm. The two-photon wavefunction is symmetrized at this wavelength, see
Fig.~\ref{fig:symmetry}.}
\end{figure}

Referring to Eq.~(\ref{pifn}), it is clear that if $D_+/D$ is made to vanish,
then $\Pi(t_+,t_-)$ becomes symmetric. Figure~\ref{fig:dpdm} shows the value of
$D_+$ and $D$ for a type-II BBO crystal as a function of pump pulse central
wavelength. Note that $D_+=0$ when the central wavelength of the pump pulse is
757 nm. The two-photon wavefunction for this case is shown in
Fig.~\ref{fig:symmetry} for a crystal thickness of 2 mm and a pump bandwidth of
8 nm.

In this example, this approach has an additional advantage: the entangled
photon pairs are emitted with central wavelengths of 1514 nm, which is within
the standard fiber communication band. Such pulsed entangled photons at
communication wavelengths may be useful for building practical quantum key
distribution systems using commercially installed optical fibers. At this
wavelength, of course, single photon detection is more problematic. However,
the recent development of single photon counting techniques using InGaAs
avalanche photodiodes may soon provide good single photon counters at this
wavelength \cite{counter}.

We have shown that by making the $D_+$ term vanish through a careful choice of
the pump wavelength, an initially asymmetric two-photon wavefunction can be
made symmetric. Since the temporal and spectral properties of the photons are
related by simple Fourier Transforms, it should come as no surprise that
$D_+=0$ leads to a spectrally symmetric state, as well.

The spectral properties of the two-photon state from ultrafast type-II SPDC are
best illustrated in plots of the two-photon joint spectrum, which can be
regarded as a probability distribution for the photon frequencies. Recall that
the two-photon state $|\psi\rangle$ given in Eq.~(\ref{spdcstate}) contains the
phase mismatch term ${\rm sinc}(\Delta L/2)$ and the pump envelope term
$\mathcal{E}_p(\omega_e+\omega_o)$. The joint spectrum function is simply the
square modulus of the product of these two terms:
$$
S(\omega_e,\omega_o)=|\text{sinc}(\Delta L/2)
\mathcal{E}_p(\omega_e+\omega_o)|^2.
$$

\begin{figure}[htb]
\includegraphics[width=3.3in]{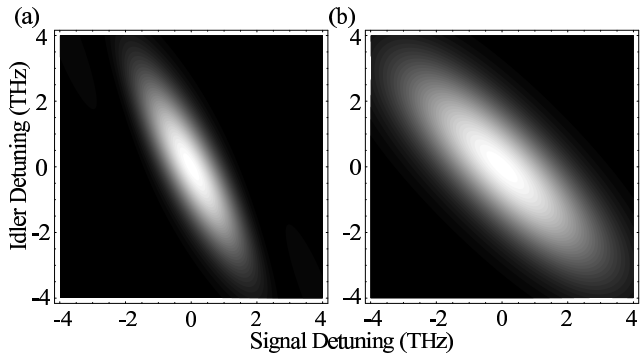}
\caption{\label{fig:anticorr} Calculated joint spectrum showing
frequency-anticorrelation between the signal and the idler photons. (a) Typical
frequency-anticorrelated state with asymmetric joint spectrum. Pump central
wavelength is 400 nm. (b) The asymmetry has now disappeared. The pump central
wavelength is 757 nm and the bandwidth is 8 nm.}
\end{figure}

The joint spectrum function for a typical ultrafast type-II SPDC configuration
(type-II BBO pumped with an ultrafast UV pulse) is shown in
Fig.~\ref{fig:anticorr}(a). Here we have assumed $L=2$ mm, the pump wavelength
is 400 nm, the pump bandwidth is 2 nm, and the SPDC wavelength is 800 nm. Note
that the joint spectrum function is asymmetric: the frequency range of the
idler photon is much larger than that of the signal. The joint spectrum becomes
symmetric, however, when $D_+=0$, as shown in Fig.~\ref{fig:anticorr}(b). Note
that the signal and the idler photons have identical spectra. Thus, the photon
pair is distinguishable only in the polarizations of the photons, as required
for a polarization entangled state.

The spectral properties of the photon pairs represented in
Fig.~\ref{fig:anticorr}(b) display the tendency of frequency-anticorrelation in
the sense that a positive detuning for one photon is accompanied by a negative
detuning for the other. This effect follows from the energy conservation
condition that constrains the SPDC process. In ultrafast type-II SPDC, the
anticorrelation is not as strong as in the cw-pumped case due to the broad
bandwidth of the pump pulse. However, the general tendency of anti-correlation,
$\omega_s=\omega_p/2 \pm \omega$ and $\omega_i=\omega_p/2 \mp \omega$ where
$\omega$ is the detuning frequency, is clearly visible in
Fig.~\ref{fig:anticorr}. This frequency-anticorrelation, however, is not a
required feature of the two-photon state. In ultrafast type-II SPDC, two-photon
states with novel spectral characteristics $-$ the frequency-correlated state
and the frequency-uncorrelated state $-$ may be generated through appropriate
choices of the parameters that affect the joint spectrum.

The idea behind two-photon states with novel spectral characteristics is not
new. The output characteristics of a beamsplitter and a Mach-Zehnder
interferometer for frequency-correlated and frequency-uncorrelated photon pairs
are theoretically studied in Ref.~\cite{campos} with no discussion of how such
states might be generated. Frequency-correlated states are also studied in
Ref.~\cite{giov}, but the discussions are limited to the generation of
frequency-correlated states using quasi-phase matching in a periodically poled
crystal. Here, we discuss how frequency-correlated states and
frequency-uncorrelated states may be generated via appropriate spectral
engineering of the two-photon state in ordinary bulk crystals. Since our main
focus is polarization entangled photon pairs with novel spectral
characteristics, we restrict our attention to the case in which the two-photon
joint spectrum is symmetric, i.e., $D_+=0$.

Frequency-correlated state generation via the SPDC process is somewhat
counter-intuitive since conservation of energy requires the frequencies of the
photon pair to sum to the frequency of the pump photon. This requirement
imposes a strong constraint in cw-pumped SPDC, where the pump the pump field is
monochromatic. In ultrafast SPDC, however, the pump field has a bandwidth of
several terahertz frequency (several nanometers in wavelength) and so the
frequencies of the photon pair need only sum to some value within the range of
pump frequencies. This extra freedom makes it possible to generate the
frequency-correlated state.

\begin{figure}[htb]
\includegraphics[width=3.3in]{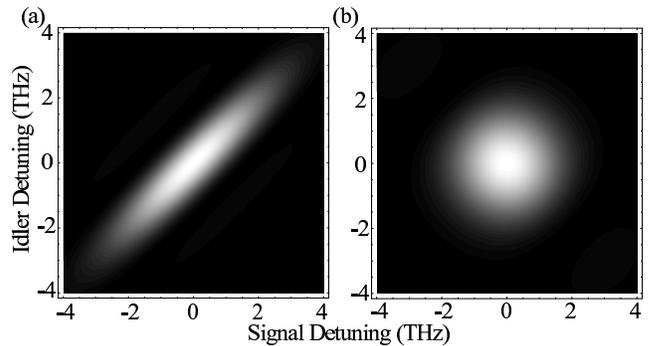}
\caption{\label{fig:corr} (a) Calculated joint spectrum showing
frequency-correlated two-photon state. The pump bandwidth is 20 nm and the
crystal is 12 mm thick type II BBO. (b) Frequency-uncorrelated state. The pump
bandwidth is 10 nm and the crystal is 5 mm thick type-II BBO. The pump central
wavelength is 757 nm for both cases to ensure $D_+=0$ is satisfied.}
\end{figure}

Examination of the joint spectrum function $S(\omega_s,\omega_i)$ suggests that
if the bandwidth of the pump envelope $\mathcal{E}_p(\omega_s+\omega_i)$ is
large enough and if the parameters are chosen such that the $\text{sin}(\Delta
L/2)$ may be approximated by the delta function $\delta(\omega_s-\omega_i)$,
then the two-photon state becomes
\begin{equation}
|\psi\rangle = C \int d\omega \,
\mathcal{E}_p(2\omega)a_s^\dagger(\omega)a_i^\dagger(\omega)|0\rangle,\label{eq:corr}
\end{equation}
where $C$ is a constant. Eq.~(\ref{eq:corr}) shows the signature of the
frequency-correlated state: $\omega_s=\omega_p/2\pm\omega$ and
$\omega_i=\omega_p/2\pm\omega$. Figure~\ref{fig:corr}(a) shows the calculated
joint spectrum function for the frequency-correlated two-photon state. In this
example, the pump central wavelength was set to 757 nm in order to satisfy the
condition $D_+=0$, i.e., to assure direct generation of polarization entangled
photon pairs. This also has the effect of properly aligning the
$\text{sinc}(\Delta L/2)$ function. The only other requirement is that this
function should be narrow enough that it may be approximated as
$\delta(\omega_s-\omega_i)$. This is achieved by increasing the value of $L$,
the crystal thickness, since the width of the sinc function is inversely
proportional to the crystal length. In Fig.~\ref{fig:corr}(a), it is assumed
that a 12 mm thick type-II BBO crystal is pumped by a 20 nm bandwidth ultrafast
pulse centered at 757 nm and that, at zero detuning, the wavelengths of the
SPDC photons are 1514 nm. The structure of frequency-correlation is clearly
illustrated: as the signal detuning increases, the idler detuning is also
increased. Such frequency-correlated states have been shown to be useful for
certain quantum metrology applications \cite{clock}.

By slightly modifying the condition for the generation of the
frequency-correlated state, it is possible to generate the
frequency-uncorrelated state. By frequency-uncorrelated state, we mean that the
frequencies of the two photons are uncorrelated, in the sense that the range of
the available frequencies for a particular photon is completely independent of
the frequency of its conjugate. As far as the photons are concerned, this means
that the spectral properties of one photon are in no way correlated with the
spectral properties of the other. The joint spectrum of such a state is shown
in Fig.~\ref{fig:corr}(b) where the joint spectrum function is calculated for
type-II SPDC in a 5-mm thick BBO crystal pumped by a 10 nm bandwidth ultrafast
pump centered at 757 nm. Again, the pump pulse central wavelength is set to 757
nm to ensure symmetry in the joint spectrum. The more general case is discussed
in Ref.~\cite{grice}, where it is also shown that such states are essential in
experiments involving interference between photons from different SPDC sources.

\section{Summary}

The generation of polarization entangled states requires the coherent
superposition of two two-photon wavefunctions. An additional requirement is
that the two wavefunctions must be identical in all respects except
polarization. This is not the case, in general, when the photon pairs originate
in a type-II SPDC process, although only a simple delay is required when the
process is pumped by a cw laser. When an ultrafast pumping scheme is employed,
however, these techniques typically do not result in the complete restoration
of quantum interference. Some improvement is seen with (i) thin nonlinear
crystals and/or (ii) narrow spectral filters before the detectors.
Unfortunately, both of these techniques result in greatly reduced count rates.
By engineering the two-photon state of ultrafast type-II SPDC in time or in
frequency, it is possible to recover the quantum interference without
discarding any photon pairs. In addition, spectral engineering holds the
promise of two-photon states with novel spectral properties which have not been
available previously.

The temporal engineering of the two-photon state is accomplished through a
novel interferometric technique which changes the temporal distribution of the
two-photon amplitude $\mathcal{A}(t_+,t_-)$, which is comprised of two
two-photon wavefunctions $\Pi(t_+,t_-)$. The shapes of the wavefunctions are
not altered in this process $-$ only the way in which they overlap in
$\mathcal{A}(t_+,t_-)$. The temporally engineered source does not suffer the
loss of quantum interference common in ultrafast type-II SPDC and the
interferometer can be viewed as a universal Bell-state synthesizer since the
quantum interference is independent of the crystal properties, the bandwidth of
spectral filters, the bandwidth of the pump laser, and the wavelengths.

A different approach is taken in the spectral engineering technique. Here, the
asymmetric two-photon wavefunction $\Pi(t_+,t_-)$ from ultrafast type-II SPDC
is symmetrized through careful control of the crystal and pump properties. Such
spectrally engineered two-photon states exhibit complete quantum interference
with no need for auxiliary interferometric techniques. The spectral engineering
techniques may also be employed in the generation of two-photon states with
novel spectral characteristics: the frequency-correlated state and the
frequency-uncorrelated state. Unlike typical two-photon states from the SPDC
process, which exhibit strong frequency-anticorrelation, the
frequency-correlated state is characterized by a strong positive frequency
correlation, while the frequencies of the two photons are completely
uncorrelated in the frequency-uncorrelated state.

The temporal and spectral engineering techniques studied in this paper are
expected play an important role in generating the pulsed entangled photon pairs
essential in applications such as practical quantum cryptography, multi-photon
entangled state generation, multi-photon interference experiments, etc. Since
temporal engineering allows one to control the decoherence in a stable way, it
also facilitates the study of the effects of decoherence in entangled
multi-qubit systems. In addition, frequency-correlated discussed in this paper
may be useful for certain quantum metrology applications.

\begin{acknowledgements}
We would like to thank M.V. Chekhova, S.P. Kulik, M.H. Rubin, and Y. Shih for
helpful discussions. This research was supported in part by the U.S. Department
of Energy, Office of Basic Energy Sciences. The Oak Ridge National Laboratory
is managed for the U.S. DOE by UT-Battelle, LLC, under contract
No.~DE-AC05-00OR22725.
\end{acknowledgements}


\begin{thebibliography}{}

\bibitem{eprb} A. Einstein, B. Podolsky, and N. Rosen, Phys. Rev.
\textbf{47}, 777 (1935); J.S. Bell, Speakable and unspeakable in quantum
mechanics, Cambridge University Press, New York (1987).

\bibitem{nielson} A. Steane, Rep. Prog. Phys. \textbf{61}, 117 (1998);
M.A. Nielson and I.L. Chuang, Quantum Computation and Quantum Information,
Cambridge University Press, New York (2000).

\bibitem{litho} A.N. Boto \textit{et al.}, Phys. Rev. Lett. \textbf{85}, 2733
(2000); M. D'Angelo, M.V. Chekhova, and Y. Shih, Phys. Rev. Lett. \textbf{87},
013602 (2001);

\bibitem{gyro} J.P. Dowling, Phys. Rev. A \textbf{57}, 4736 (1998).

\bibitem{clock} R. Jozsa \textit{et al.}, Phys. Rev. Lett. \textbf{85}, 2010
(2000); V. Giovannetti, S. Lloyd, and L. Maccone, Phys. Rev. A \textbf{65},
022309 (2002).

\bibitem{klyshkobook} D.N. Klyshko, Photons and Nonlinear Optics, Gordon and
Breach, New York (1988).

\bibitem{kiess} T.E. Kiess \textit{et al.}, Phys. Rev. Lett. \textbf{71}, 3893
(1993).

\bibitem{shih1} Y.H. Shih and A.V. Sergienko, Phys. Lett. A \textbf{186}, 29
(1994); Y.H. Shih and A.V. Sergienko, Phys. Lett. A \textbf{191}, 201 (1994)

\bibitem{rubin} M.H. Rubin \textit{et al.},
Phys. Rev. A \textbf{50}, 5122 (1994).

\bibitem{kwiat1} P.G. Kwiat \textit{et al.}, Phys. Rev. Lett. \textbf{75}, 4337
(1995).

\bibitem{keller} T.E. Keller \textit{et al.}, Phys. Rev. A \textbf{57}, 2076
(1998).

\bibitem{decaro} L. De Caro and A. Garuccio, Phys. Rev. A \textbf{50}, R2803
(1994); S. Popescu, L. Hardy, and M. Zukowski, Phys. Rev. A \textbf{56}, R4353
(1997); M. Zukowski \textit{et al.}, Phys. Rev. A \textbf{60}, R2614 (1999).

\bibitem{pulsedspdctheory} T.E. Keller and M.H. Rubin, Phys. Rev. A
\textbf{56}, 1534 (1997); W.P. Grice and I.A. Walmsley, Phys. Rev. A
\textbf{56}, 1627 (1997).

\bibitem{pulsedspdcexp} G. Di Giuseppe \textit{et al.}, Phys. Rev. A 56, R21
(1997); W.P. Grice \textit{et al.}, Phys. Rev. A \textbf{57}, R2289 (1998);
Y.-H. Kim \textit{et al.}, Phys. Rev. A \textbf{64}, 011801(R) (2001); Y.-H.
Kim \textit{et al.}, Phys. Rev. Lett. \textbf{86}, 4710 (2001).

\bibitem{note} The common misconception is that the photon pairs emerging
from these two cross sections are automatically polarization entangled. As we
have discussed in section \ref{type2spdc}, the polarization state of the photon
pair is in a mixed state
$$
\rho_{mix} = \frac{1}{2}\left(|H_1\rangle|V_2\rangle\langle V_2|\langle H_1| +
|V_1\rangle|H_2\rangle\langle H_2|\langle V_1| \right).
$$

\bibitem{kim3} D. Branning \textit{et al.}, Phys. Rev. Lett. \textbf{83},
955 (1999); J.H. Shapiro and F.N.C. Wang, J. Opt. B \textbf{2}, L1 (2000);
Y.-H. Kim \textit{et al.}, Phys. Rev. A \textbf{63}, 062301 (2001); A.V.
Burlakov \textit{et al.}, \textit{ibid.} \textbf{64}, 041803(R) (2001).


\bibitem{werner} R.F. Werner, Phys. Rev. A \textbf{40}, 4277 (1989);
P.G. Kwiat \textit{et al.}, Nature \textbf{409}, 1014 (2001); A.G. White
\textit{et al.}, Phys. Rev. A \textbf{65}, 012301 (2001).

\bibitem{grice} W.P. Grice, A.B. U'Ren, and I.A. Walmsley, Phys. Rev. A \textbf{64},
063815 (2001).

\bibitem{counter} A. Lacaita \textit{et al.}, Applied Optics \textbf{35}, 2986
(1996); G. Ribordy \textit{et al.}, Applied Optics \textbf{37}, 2272 (1998); A.
Karlsson \textit{et al.}, IEEE Circuits \& Devices, Nov., 34, (1999); P.A.
Hiskett \textit{et al.}, Applied Optics \textbf{39}, 6818 (2000).

\bibitem{campos} R.A. Campos, B.E.A. Saleh, and M.C. Teich, Phys. Rev. A \textbf{42}, 4127
(1990).

\bibitem{giov} V. Giovannetti \textit{et al.}, quant-ph/0109135.

\end{thebibliography}
\end{document}